\documentclass[preprintnumbers,amssymb,amsmath,superscriptaddress,nofootinbib]{revtex4}
\usepackage{graphicx}
\usepackage{dcolumn}
\usepackage{bm}
\usepackage{comment}
\usepackage{color}
\definecolor{red}{rgb}{1.0,0.0,0.0}
\definecolor{green}{rgb}{0.0,1.0,0.0}
\definecolor{blue}{rgb}{0.0,0.0,1.0}
\definecolor{violet}{rgb}{1.0,0.0,1.0}
\definecolor{black}{rgb}{0.0,0.0,0.0}
\definecolor{darkgray}{rgb}{0.33,0.33,0.33}
\definecolor{lightgray}{rgb}{0.66,0.66,0.66}
\begin{document} 
\input epsf.tex
\newcommand{\beq}{\begin{eqnarray}}
\newcommand{\eeq}{\end{eqnarray}}
\newcommand{\nn}{\nonumber}
\def\ltap{\ \raise.3ex\hbox{$<$\kern-.75em\lower1ex\hbox{$\sim$}}\ }
\def\gtap{\ \raise.3ex\hbox{$>$\kern-.75em\lower1ex\hbox{$\sim$}}\ }
\def\CO{{\cal O}}
\def\CL{{\cal L}}
\def\CM{{\cal M}}
\def\tr{{\rm\ Tr}}
\def\CO{{\cal O}}
\def\CL{{\cal L}}
\def\CM{{\cal M}}
\def\mpl{M_{\rm Pl}}
\newcommand{\bel}[1]{\be\label{#1}}
\def\al{\alpha}
\def\bt{\beta}
\def\eps{\epsilon}
\def\eg{{\it e.g.}}
\def\ie{{\it i.e.}}
\def\mn{{\mu\nu}}
\newcommand{\rep}[1]{{\bf #1}}
\def\be{\begin{equation}}
\def\ee{\end{equation}}
\def\bea{\begin{eqnarray}}
\def\eea{\end{eqnarray}}
\newcommand{\eref}[1]{(\ref{#1})}
\newcommand{\Eref}[1]{Eq.~(\ref{#1})}
\newcommand{\gsim}{ \mathop{}_{\textstyle \sim}^{\textstyle >} }
\newcommand{\lsim}{ \mathop{}_{\textstyle \sim}^{\textstyle <} }
\newcommand{\vev}[1]{ \left\langle {#1} \right\rangle }
\newcommand{\bra}[1]{ \langle {#1} | }
\newcommand{\ket}[1]{ | {#1} \rangle }
\newcommand{\ev}{{\rm eV}}
\newcommand{\kev}{{\rm keV}}
\newcommand{\Mev}{{\rm MeV}}
\newcommand{\gev}{{\rm GeV}}
\newcommand{\tev}{{\rm TeV}}
\newcommand{\mev}{{\rm MeV}}
\newcommand{\meV}{{\rm meV}}
\newcommand{\mnu}{\ensuremath{m_\nu}}
\newcommand{\nnu}{\ensuremath{n_\nu}}
\newcommand{\mlr}{\ensuremath{m_{lr}}}
\newcommand{\acc}{\ensuremath{{\cal A}}}
\newcommand{\mav}{MaVaNs}
\newcommand{\disc}[1]{{\bf #1}} 
\newcommand{\mh}{{m_h}}
\newcommand{\hb}{{\cal \bar H}}
\newcommand{\me}{\mbox{${\rm \not\! E}$}}
\newcommand{\met}{\mbox{${\rm \not\! E}_{\rm T}$}}

\renewcommand{\thefootnote}{\fnsymbol{footnote}}

\title{Atmospheric Tau Neutrinos in a Multi-kiloton Liquid Argon Detector}
\author{Janet Conrad}
\affiliation{Massachusetts Institute of Technology, Cambridge, MA 02139, USA}
\author{Andr\'e de Gouv\^ea}
\affiliation{Department of Physics \& Astronomy, Northwestern University, IL 60208-3112, USA}
\author{Shashank Shalgar}
\affiliation{Department of Physics \& Astronomy, Northwestern University, IL 60208-3112, USA}
\author{Joshua Spitz}
\affiliation{Department of Physics, Yale University, New Haven, CT 06520, USA}
\preprint{NUHEP-TH/10-11}
\date{\today}
\begin{abstract}
An ultra-large Liquid Argon Time Projection Chamber-based neutrino detector will have the uncommon ability
to detect atmospheric $\nu_\tau$/$\overline{\nu}_{\tau}$ events.    This paper discusses the most promising modes for identifying
    charged current $\nu_\tau$/$\overline{\nu}_{\tau}$, and shows that, with simple kinematic cuts,  $\sim$30 $\nu_\tau+$$\overline{\nu}_{\tau}$ interactions
    can be isolated in a 100~kt$\cdot$yr exposure, with greater than $4\sigma$ significance. This sample is sufficient to perform flux-averaged total cross-section and cross-section shape parameterization measurements -- the first steps toward using $\nu_\tau$/$\overline{\nu}_{\tau}$ to search for physics beyond the Standard
    Model.

\end{abstract}

\maketitle

\setcounter{equation}{0} \setcounter{footnote}{0}
\section{Introduction}
\label{sec:Intro}




Of all observed Standard Model particles, we have the least direct experimental knowledge of the tau neutrino, $\nu_\tau$/$\overline{\nu}_{\tau}$\footnote{The distinction between $\nu_\tau$ and $\overline{\nu}_{\tau}$ is made now in order to avoid ambiguity later in the paper.}. A high statistics, dedicated experiment sensitive to $\nu_{\tau}$/$\overline{\nu}_{\tau}$ Charged Current (CC) interactions has never been performed and would add significantly to our understanding of electroweak interactions and be sensitive to certain hard-to-get manifestations of new physics~\cite{Adams:2009mc}.  Indeed, any opportunity to collect a significant sample of CC~$\nu_\tau$/$\overline{\nu}_{\tau}$ interactions and simply measure the cross-section and perform basic tests of Standard Model predictions would qualitatively improve our knowledge of the third neutrino weak eigenstate.

The path to the observation of the $\nu_{\tau}$ was long, not unlike that of the electron neutrino, $\nu_e$, and the muon neutrino, $\nu_{\mu}$. The discovery of the tau-lepton in 1975~\cite{Perl:1975bf} led to the assumption of a third neutrino, $\nu_{\tau}$, the weak-isospin partner of the third charged lepton, $\tau$. Since then, indirect information on the $\nu_\tau$ has been collected from a wide
range of tau decay analyses~\cite{Abe:1997dy,Barate:1997zg,Buskulic:1993mn,Acciarri:1996ht,Chapkin:1998pu,Ackerstaff:1998es,Abashian:2000cx}. The fact that the $\nu_{\tau}$ is a state orthogonal to $\nu_{\mu}$ and $\nu_e$, for example, was first indirectly revealed by the LEP experiments via precision measurements of the $Z$-boson width~\cite{LEP:2005ema}.  The direct observation of CC~$\nu_\tau$/$\overline{\nu}_{\tau}$ interactions is a very recent (21$^{\rm st}$ century) development. The first events were presented by the DONuT experiment in 2000 and published in 2001~\cite{Kodama:2000mp}.  To date, ten CC~$\nu_\tau$ interactions have been observed, nine by DONuT and one by OPERA~\cite{Kodama:2008zz,Agafonova:2010dc}\footnote{This is dwarfed by, for example, the world's growing sample of reconstructed top quark events, which currently consists of well over 1000 events~\cite{Abazov:2009ae,Aaltonen:2010pe}.}. DONuT was a short-baseline, emulsion-based experiment.  The $\nu_\tau$/$\overline{\nu}_{\tau}$s were produced by a fixed target 800~GeV proton beam configuration,
primarily through the decay of $D_s$ mesons, with the relevant branching ratio ${\cal B}(D_s^- \to \tau^- \overline{\nu}_\tau) = (6.6 \pm 0.6)$\%~\cite{Amsler:2008zzb}. The $\nu_\tau$ and $\overline{\nu}_\tau$ contributed about 3\% of the total neutrino flux. 

An alternative method for producing/detecting $\nu_\tau$/$\overline{\nu}_{\tau}$ relies on neutrino oscillations. Our current understanding of the neutrino oscillation data~\cite{Schwetz:2008er,GonzalezGarcia:2010er} indicates that neutrinos produced as $\nu_\mu$/$\overline{\nu}_{\mu}$s will oscillate mostly into $\nu_{\tau}$/$\overline{\nu}_{\tau}$s with an oscillation frequency related to the largest (in magnitude) of the two independent mass squared-differences, $|\Delta m^2_{13}|\sim2\times 10^{-3}$~eV$^2$. This is true as long as the oscillation length associated with the smallest mass-squared difference, $\Delta m^2_{12}\sim 8\times 10^{-5}$~eV$^2$, is much longer than the characteristic baseline of the experiment. The relevant mixing angle is consistent with maximal ($\sin^22\theta_{23}\sim 1$)~\cite{Schwetz:2008er,GonzalezGarcia:2010er} so a detector placed at one of the oscillation maxima and exposed to an originally-$\nu_\mu$/$\overline{\nu}_{\mu}$ beam provides an ideal setup for collecting a large sample of CC~$\nu_\tau$/$\overline{\nu}_{\tau}$ interactions.  In practice, the design of such an experiment has been demonstrated to be challenging. A large oscillation phase demands baselines above (and, preferably, well above) 1000~km, as the tau production threshold requires $\nu_{\tau}$/$\overline{\nu}_{\tau}$s with laboratory energies above 3.5~GeV. To date, it has not been possible to produce a beam with an experimentally significant neutrino flux at distances beyond about 1000~km~\cite{Kopp:2006ky,Barger:2007yw}. The OPERA experiment~\cite{Mauri:2009zz,Pessard:2009vc}, a 1.25~kt emulsion-based detector, is aimed at directly observing CC~$\nu_\tau$ç events in a long baseline beam. Unfortunately, the $L/E\sim 700~{\rm km}/20~{\rm GeV}$ factor does not allow for the collection of a large data sample. As of this writing, one $\nu_{\tau}$ candidate event has been observed~\cite{Agafonova:2010dc}, with 10.4 events expected after five years of running at design luminosity~\cite{Mauri:2009zz}.

Emulsion detectors provide the strongest resolving power for CC~$\nu_\tau$/$\overline{\nu}_{\tau}$ interactions. Such detectors isolate the events
through the observation of a ``kink" from the short-lived tau decay.  However, automatic scanning of the emulsion is a time-intensive process which cannot proceed in real time.  An alternative method, with real-time, fast event reconstruction using drift chambers was employed by the NOMAD experiment in an oscillation search~\cite{Astier:2001yj}.  No $\nu_\tau$ interactions were observed as the accessible $\Delta m^2$ range was outside of what we now know is allowed. Liquid Argon Time Projection Chamber (LArTPC) detectors have been proposed for $\nu_\tau$/$\overline{\nu}_{\tau}$ event searches, but have yet to be specifically employed for this purpose~\cite{Adams:2009mc}. Very recently, the ICARUS T-600 neutrino detector started operating in Gran Sasso. The ICARUS T-600 is a 600~ton LArTPC exposed, like OPERA, to the CNGS beam~\cite{icarus,Arneodo:2001tx}. Although the experiment's main goal is to observe $\nu_{\mu}$ disappearance, ICARUS is also equipped to observe a number of CC~$\nu_{\tau}$ interactions~\cite{Dracos:2010zz}. Unfortunately, the expected event sample is too small to allow one to perform a cross-section measurement. In both the NOMAD and ICARUS experiments, the beam direction is used to search for evidence of missing transverse momentum consistent with tau production and decay.  

The proposed 20~kt LArTPC at the Deep Underground Science and Engineering Laboratory (DUSEL) opens a new opportunity to collect a significant sample of $\nu_\tau$/$\overline{\nu}_{\tau}$ events. Although the detector will be significantly larger than OPERA, the combination of longer baseline and significantly lower beam energy leads to anywhere from 0.2-4.0 expected CC~$\nu_\tau$ beam-oscillated events/MW$\cdot10^{7}$s/kt, depending on the (as-yet-undecided) beam tune\footnote{We do not speculate on the possibility of LBNE-beam-based $\nu_\tau$/$\overline{\nu}_{\tau}$ detection as it is extremely dependent on the largely undecided beam parameters (e.g. beam energy spectrum).}\cite{marybishai}. However, as a deep underground detector, this experiment will be sensitive to atmospheric neutrino interactions. A natural sample of high energy, earth-diameter-as-baseline oscillations, has the drawback of having no clear beam direction.  In this paper, however, we show that this problem can be overcome so that a significant number of atmospheric $\nu_\tau$/$\overline{\nu}_{\tau}$ events can be identified.   We present a discussion of the capability of $\nu_\tau$+$\overline{\nu}_{\tau}$ cross-section measurements an an example of the physics potential of this signal. 

The idea of a search for atmospheric $\nu_\tau$/$\overline{\nu}_{\tau}$ appearance has been pursued before. The 22.5~kt (fiducial) Super-Kamiokande detector~\cite{Abe:2006fu} results disfavor the no $\nu_\tau$/$\overline{\nu}_{\tau}$ appearance hypothesis at the 2.4$\sigma$ level.  However, this analysis was hampered by the lack of precision track reconstruction and particle identification inherent to Cherenkov-based detectors. A multi-kiloton LArTPC, on the other-hand, will provide well-reconstructed events which can be used to isolate the $\nu_\tau$/$\overline{\nu}_{\tau}$ interaction signal in a very convincing way, as we present below. We note that the authors of~\cite{Giordano:2010pr} and~\cite{icanoe,icanoe2} have also considered the possibility of studying atmospheric $\nu_\tau$/$\overline{\nu}_{\tau}$ using the Ice Cube Deep Core Array and a LArTPC (with magnetic calorimeter), respectively.
The physics issues involved in the calculation of the $\nu_{\tau}$/$\overline{\nu}_{\tau}$ cross-section at high energies relevant to Ice Cube Deep Core Array-like experiments have recently been discussed in the literature~\cite{Jeong:2010nt}.

This paper is organized as follows. In Section~\ref{sec:physics}, we discuss the ``production'' of $\nu_\tau$/$\overline{\nu}_{\tau}$ via atmospheric neutrino oscillations and review the cross-section and kinematics of  CC~$\nu_{\tau}$/$\overline{\nu}_{\tau}$ interactions. In Section~\ref{sec:experiment}, we present a brief overview of LArTPC technology and details of our CC~$\nu_{\tau}$/$\overline{\nu}_{\tau}$ event simulation and the relevant CC and Neutral Current (NC) atmospheric neutrino induced backgrounds. In Sections~\ref{sec:tau_to_x}, \ref{sec:tau_to_n_pi}, and \ref{sec:tau_to_ell}, we discuss the search for CC~$\nu_{\tau}$/$\overline{\nu}_{\tau}$ events via statistical inference in various tau decay modes, and offer some concluding remarks in Section~\ref{sec:conclusion}.  


\setcounter{equation}{0}
\section{Atmospheric Tau Neutrinos: Production and Detection}
\label{sec:physics}

In order to understand an ultra-large LArTPC's ability to detect $\nu_\tau$/$\overline{\nu}_{\tau}$, it is important to know the properties of the atmospheric $\nu_{\tau}$/$\overline{\nu}_{\tau}$ ``beam'' and the Standard Model expectation for the CC $\nu_{\tau}N\to\tau X$ cross-section, where $N$ is a nucleon and $X$ is any hadronic final state. Both issues are discussed in this section.

\subsection{The Atmospheric $\nu_{\tau}$/$\overline{\nu}_{\tau}$ Flux}


Most atmospheric neutrinos are a result of pion decays (with a sub-leading kaon component), with the pions produced when cosmic rays interact with the Earth's atmosphere several kilometers above sea level. One naively expects a $\nu_{\mu}$/$\overline{\nu}_{\mu}$ to $\nu_e$/$\overline{\nu}_{e}$ production ratio close to two with virtually zero $\nu_{\tau}$/$\overline{\nu}_{\tau}$s\footnote{A very small $\nu_{\tau}$/$\overline{\nu}_{\tau}$ component to the parent atmospheric neutrino flux is expected from, for example, charm production. 
This addition is both very small and of higher energy than the atmospheric neutrinos considered here, and will be neglected henceforth~\cite{Pasquali:1998xf,Bulmahn:2010pg}.}.

\begin{figure}
\hspace{-1.5cm}
\centerline{
\includegraphics[width=16cm]{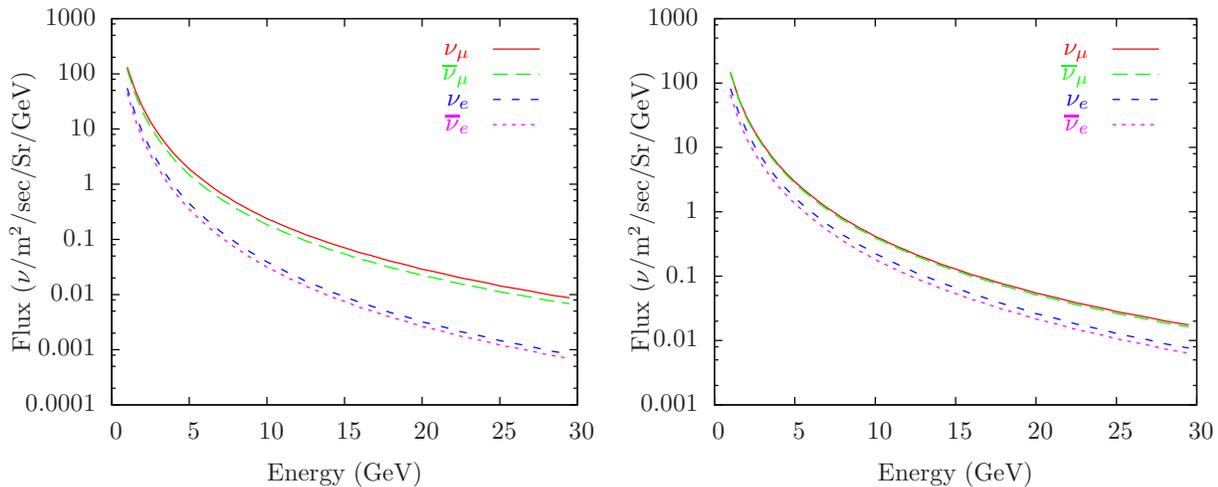}
}
 \caption{Atmospheric neutrino flux as a function of the neutrino energy, for $\cos\theta_{\mathrm{zenith}}=-1$ (left) and 0 (right).}
\label{fig:flux_energy}
\end{figure}

\begin{figure}
\hspace{-1.5cm}
\centerline{
\includegraphics[width=16cm]{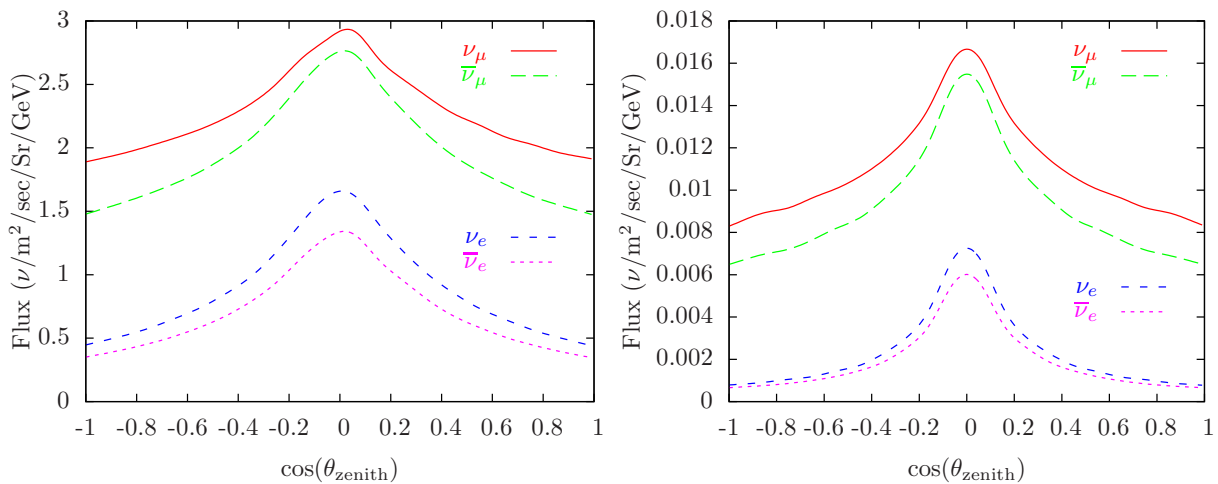}
}
\caption{Atmospheric neutrino flux as a function of $\cos\theta_{\mathrm{zenith}}$ for $E=5$~GeV (left) and $30$~GeV (right).}
\label{fig:flux_angle}
\end{figure}

Figure~\ref{fig:flux_energy} depicts the flux of $\nu_{\mu}$, $\overline{\nu}_{\mu}$, $\nu_e$, and $\overline{\nu}_e$ as a function of the neutrino energy, for $\cos\theta_{\mathrm{zenith}}=-1$ (left) and $\cos\theta_{\mathrm{zenith}}=0$ (right), where $\cos\theta_{\mathrm{zenith}}$ is the cosine of the zenith angle\footnote{$\cos\theta_{\mathrm{zenith}}=+1$ corresponds to neutrinos coming from above and $\cos\theta_{\mathrm{zenith}}=-1$ to neutrinos coming from the antipodal point on the Earth.}, according to~\cite{Honda:2004yz}. Figure~\ref{fig:flux_angle} depicts the fluxes as a function of zenith angle for $E=5$~GeV (left) and $E=30$~GeV (right). The uncertainty on the absolute normalization of the atmospheric neutrino fluxes is estimated to be around 20\%, while the energy and zenith angle dependencies (shapes) are known to within 5\%~\cite{Fogli:2002pt,GonzalezGarcia:2007ib}.

Between production and detection (while the neutrinos traverse a distance $L$), neutrinos oscillate. The relevant vacuum oscillation phases are,
\begin{equation}
\Phi_{ij}=\frac{|\Delta m^2_{ij}|L}{4E}=7.8\left(\frac{|\Delta m^2_{ij}|}{2.4\times 10^{-3}~\rm eV^2}\right)\left(\frac{5~\rm GeV}{E}\right)\left(\frac{L}{12756~{\rm km}}\right),
\end{equation}
where 12756~km is the maximum diameter of the Earth. $\Phi_{13}$ can be of order one for atmospheric neutrinos as long as $L$ is not much smaller than the Earth's radius.  For $\cos\theta_{\mathrm{zenith}}=0$, $L\sim\sqrt{2R_{\oplus}h}\sim 440$~km ($R_{\oplus}$ is the Earth's radius and $h\sim 15$~km is the average height above sea-level where neutrinos are produced), and $\Phi_{13}\sim0.3$ for 5~GeV neutrinos.  $\Phi_{12}$ is always small unless the neutrino energies are below 1~GeV; $\Phi_{12}<0.35$ for neutrino energies above 3.5~GeV (the CC~$\nu_{\tau}$/$\overline{\nu}_{\tau}$ production threshold) using $\Delta m^2_{12}=7.6\times 10^{-5}$~eV$^2$. In summary, the oscillated $\nu_{\tau}$/$\overline{\nu}_{\tau}$ flux above tau threshold comes from negative $\cos\theta_{\mathrm{zenith}}$ and from the dominant ``atmospheric'' oscillation frequency, proportional to $|\Delta m^2_{13}|$. 

\begin{figure}
\hspace{-1.5cm}
\centerline{
\includegraphics[width=16cm]{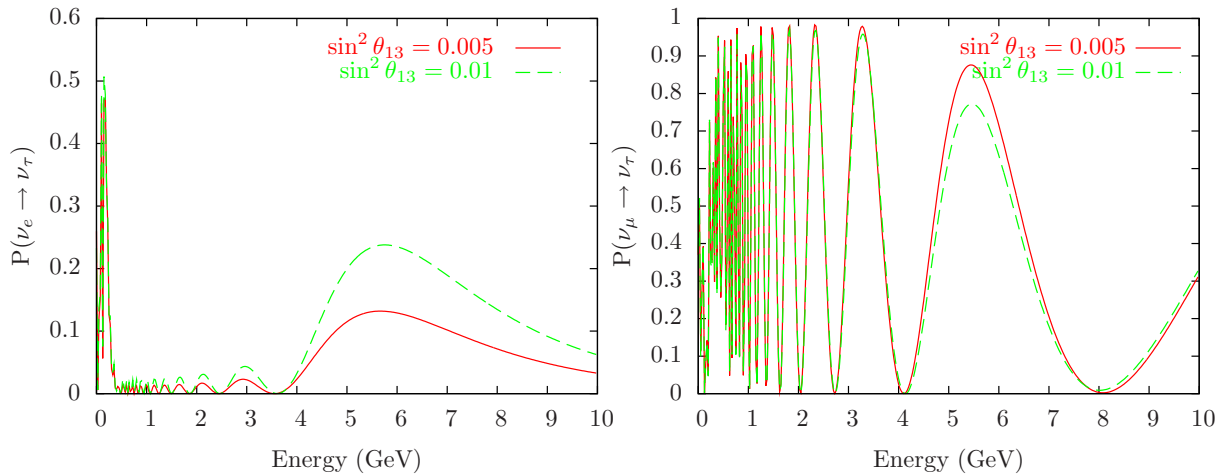}
}
\caption{$P(\nu_{e}\rightarrow \nu_{\tau})$ (left) and $P(\nu_{\mu}\rightarrow \nu_{\tau})$ (right) with $L=8000$~km for $\sin^{2}\theta_{13}=0.005$ ({\color{red}{solid}}) and 0.01 ({\color{green}{dashed}}), and a normal neutrino mass hierarchy.}
\label{fig:probs}
\end{figure}

The relevant oscillation probabilities are, ignoring matter effects, 
\begin{eqnarray}
P_{\mu\tau}&\sim\cos^2{\theta_{13}}\sin^22\theta_{23}\sin^2\Phi_{13} & \sim \sin^2\Phi_{13}, \label{eq:pmutau} \\
P_{e\tau}&\sim\sin^2{\theta_{23}}\sin^22\theta_{13}\sin^2\Phi_{13} & <  0.09\sin^2\Phi_{13}, \label{eq:petau}
 \end{eqnarray}
making use of the upper bound on $\theta_{13}$ from~\cite{GonzalezGarcia:2010er}. Figure~\ref{fig:probs} depicts $P_{e\tau}$ (left) and $P_{\mu\tau}$ (right) as a function of the neutrino energy with $L=8000$~km ($\cos\theta_{\mathrm{zenith}}=-0.63$) for different values of $\theta_{13}$ and assuming that the neutrino mass-hierarchy is normal ($\Delta m^2_{13}>0$). Note that the simplifying approximations, allowing us to write  Eqs.~(\ref{eq:pmutau}) and (\ref{eq:petau}) above, were not made in generating Figure~\ref{fig:probs}, and that matter effects were properly taken into account. The PREM density profile was used to model the density of the Earth~\cite{Dziewonski:1981xy}. 
We safely conclude, further remembering that the $\nu_{\mu}$/$\overline{\nu}_{\mu}$ flux is larger than the $\nu_e$/$\overline{\nu}_{e}$ flux, that the majority ($>$$90$\%) of the $\nu_{\tau}$/$\overline{\nu}_{\tau}$ atmospheric neutrino flux at energies above a few GeV comes from $\nu_{\mu}\to\nu_{\tau}$ ($\overline{\nu}_{\mu}\to\overline{\nu}_{\tau}$) oscillations. For large values of $\theta_{13}$ and a normal (inverted) mass hierarchy, there is a small but potentially significant fraction of $\nu_{\tau}$ ($\overline{\nu}_{\tau}$) from $\nu_e$ ($\overline{\nu}_e$) oscillations.

\begin{figure}
\hspace{-1.5cm}
\centerline{
\includegraphics[width=16cm]{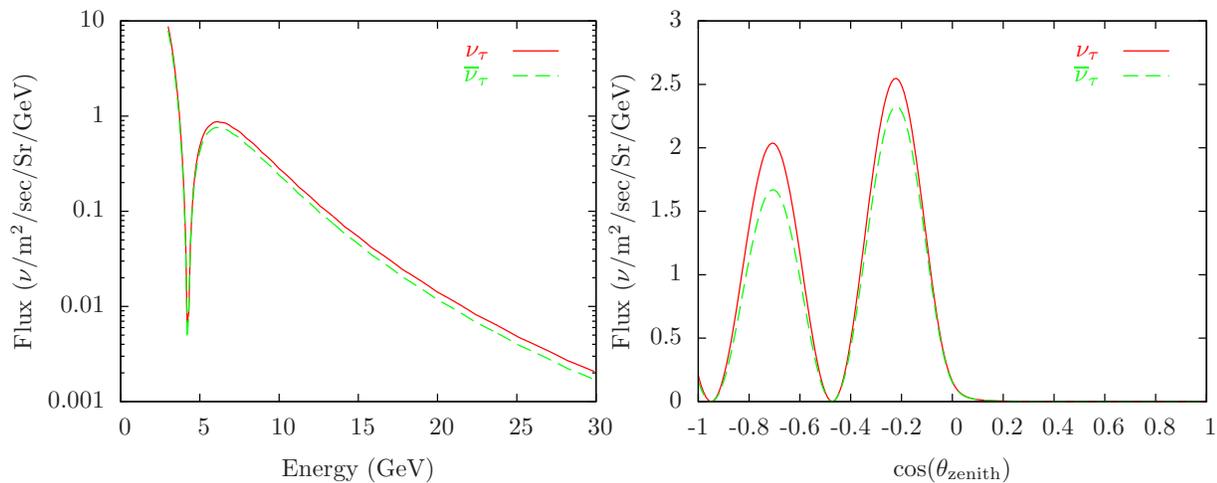}
}
\caption{Atmospheric $\nu_{\tau}$ and $\overline{\nu}_{\tau}$ fluxes from oscillations as a function of neutrino energy (left) and cosine of the zenith angle $\cos\theta_{\mathrm{zenith}}$ (right). In the left panel, the zenith angle is fixed at $\cos\theta_{\mathrm{zenith}}=-0.4$, while in the right panel the energy is fixed at $E=5$~GeV. We assume a normal neutrino mass hierarchy and $\theta_{13}=0$.}
\label{fig:nutau_flux}
\end{figure}

Figure~\ref{fig:nutau_flux} depicts the atmospheric $\nu_{\tau}$/$\overline{\nu}_{\tau}$ flux at the detector site as a function of energy (left) and zenith angle (right), for a fixed zenith angle and energy, respectively. 
We use the current best fit values of the neutrino oscillation parameters -- $\Delta m^{2}_{12}=7.6 \times 10^{-5}$ eV$^{2}$, $\Delta m^{2}_{13}=2.4 \times 10^{-3}$ eV$^{2}$, $\sin^{2}\theta_{12}=0.31$, and $\sin^{2}\theta_{23}=0.5$~\cite{Gando:2002ub,Aharmim:2008kc,Ashie:2004mr,Adamson:2008zt} -- and assume $\sin^{2}\theta_{13}=0$ and a normal neutrino mass hierarchy.
Prominent oscillatory features are clear in both panels of Figure~\ref{fig:nutau_flux}. In summary, $\nu_\tau$/$\overline{\nu}_{\tau}$s arrive at the detector mostly from ``below'' and are overwhelmingly low-energy. Also, we expect most of the CC~$\nu_{\tau}$/$\overline{\nu}_{\tau}$ initiated events to occur close to tau production threshold. 

\subsection{The $\nu_{\tau}N\to \tau X$ Cross-Section}

In the energy region of interest, CC~$\nu_{\tau}$/$\overline{\nu}_{\tau}$ interactions are well described as if the $\nu_{\tau}$s were scattering mostly off of nucleons. The kinematics of $\nu_{\tau}n\to \tau^- p$ (or $\overline{\nu}_{\tau}p\to \tau^+ n$) dictate that the minimum neutrino energy in the rest-frame of the target nucleon is
\begin{equation}
E_{\rm min}=m_{\tau}\left[1+\frac{m_{\tau}}{2m_N}\right]=3.46~{\rm GeV},
\end{equation}
where $m_{\tau}$ is the tau-lepton mass and $m_N$ is the nucleon mass, and assuming that the neutron and proton masses are the same.

Similar to CC~$\nu_e$/$\overline{\nu}_{e}$ and $\nu_{\mu}$/$\overline{\nu}_{\mu}$ scattering, CC~$\nu_{\tau}$/$\overline{\nu}_{\tau}$ scattering receives contributions from a variety of processes. At low energies, quasi-elastic (QE) scattering $\nu_{\tau}n\to \tau^- p$ ($\overline{\nu}_{\tau}p\to \tau^+ n$) dominates, while at high energies the deep-inelastic scattering (DIS) contribution is  largest. At intermediate energies, it is expected that resonance-production and other non-perturbative QCD phenomena dominate. The heavy tau mass ($m_{\tau}=1.777$~GeV) shifts the regions where the different contributions dominate towards higher energies. For example, the production of a $\Delta$-resonance requires $E_{\nu_{\tau}}\gtrsim4$~GeV for CC~$\nu_{\tau}$ scattering, as opposed to $E_{\nu_{\mu}}\gtrsim0.44$~GeV in the case of CC~$\nu_{\mu}$ scattering. For CC~$\nu_{\tau}$ scattering in the region of interest, $E_{\rm min}<E_{\nu}\lesssim 20$~GeV, the three different contributions are similar. QE is expected to dominate very close to threshold while DIS is dominant above 10~GeV or so~\cite{Paschos:2001np}.

 We will not discuss the challenges associated with computing the CC~$\nu_{\tau}$/$\overline{\nu}_{\tau}$ cross-section. 
 Instead, we will compare different results in the literature in order to illustrate the uncertainty of the situation. While a significant improvement is expected in the near future from upcoming experimental data on $\nu_{\mu}$/$\overline{\nu}_{\mu}$ and $\nu_e$/$\overline{\nu}_{e}$-scattering, a large uncertainty will remain in the $\nu_{\tau}$/$\overline{\nu}_{\tau}$ sector until more data becomes available.

\begin{figure}
\includegraphics{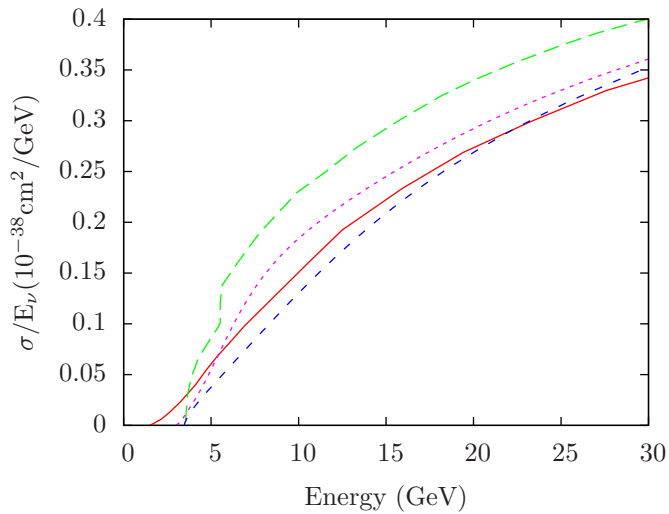}
\caption{The CC~$\nu_{\tau}$ cross section for an isoscalar target calculated according to different models: Paschos et al. ({\color{green}{long-dashed}}), Kretzer et al. ({\color{red}{solid}}) and Hagiwara et al. ({\color{blue}{short-dashed}}). The Nuance ({\color{violet}{dotted}}) cross section prediction for an argon target is also shown.} \label{fig:x-section}
\end{figure}

Table~\ref{eventnumber} contains the expected number of atmospheric CC~$\nu_\tau$/$\overline{\nu}_{\tau}$ events/100~kt$\cdot$yr from various cross-section computations found in the literature~\cite{Naples:2003ne,Kretzer:2002fr,Hagiwara:2003di,Paschos:2001np} and the Nuance~\cite{nuance} prediction. The NuTeV collaboration has extracted Fe structure functions from the $\nu$-Fe\ and $\overline{\nu}$-Fe deep inelastic differential cross-sections using its high energy sign selected beam~\cite{Naples:2003ne}. We compare the event rate from this data with those obtained from other cross-section models. The calculations done by Paschos et al.~\cite{Paschos:2001np} use leading order (LO), whereas Kretzer et al.~\cite{Kretzer:2002fr} and Hagiwara et al.~\cite{Hagiwara:2003di} use next to leading order (NLO) structure functions to calculate the total cross-section. The Kretzer et al. model takes into account charm production, tau threshold, and target mass effects. The Hagiwara et al. model accounts for the effect of polarization on the DIS cross-section, unlike other models. All of these effects are only important in the low energy region, however. The significant discrepancy in the high energy region points to the lack of data presently available.
Figure~\ref{fig:x-section} depicts different predictions for the CC~$\nu_{\tau}$ cross-section at neutrino energies below 30~GeV, where we expect the vast majority of events to occur. We bring attention to the fact that the different computations in the literature find different effective thresholds and that the shape of the rise in the cross-section for energies close to threshold also changes from one estimate to the other. We should note that the Monte Carlo model used by Super-Kamiokande~\cite{Abe:2006fu} has a cross-section very close to that of Nuance, the neutrino event generator used in this paper and discussed later.

\begin{table}
\begin{tabular}{|c|c|c|c|}
\hline
Model & $\nu_\tau$ events & $\overline{\nu}_{\tau}$ events & Total \\
\hline
NuTeV \cite{Naples:2003ne} & 56.9 & 24.9 & 81.7 \\
Kretzer et al. \cite{Kretzer:2002fr} & 37.7 & 17.5 & 55.2\\
Hagiwara et al. \cite{Hagiwara:2003di} & 33.7 & 18.5 & 52.3 \\
Paschos et al. \cite{Paschos:2001np} & 65.2 & 29.6 & 94.8\\
Nuance \cite{nuance} & 54.1 & 23.1 & 77.2\\
\hline
\end{tabular}
\caption{The expected number of CC~$\nu_\tau$/$\overline{\nu}_{\tau}$ events/100~kt$\cdot$yr on an isoscalar target as predicted by various cross-section models. The Nuance prediction for interactions on an argon target is also shown.}
\label{eventnumber}
\end{table}

In an attempt to quantify these differences, we simply parameterize the cross-section using
\begin{eqnarray}
\frac{\sigma(E)}{E}=\left(1-\exp \left(-a\sqrt{E-b}\right)\right) \times 10^{-38} \frac{\mathrm{cm}^{2}}{\mathrm{GeV}}\label{aandb},
\end{eqnarray}
where the parameter $b$ fixes the production threshold value in the lab frame and the parameter $a$ fixes the rate of increase near the threshold. The energy $E$ and parameter $b$ are in GeV. We find that this parametrization provides an excellent fit to the curves shown in Figure~\ref{fig:x-section}. The best fit values for these parameters are tabulated in~Table~\ref{parameters}. Later, we calculate the experimental sensitivity to these parameters, assuming a $\nu_{\tau}$/$\overline{\nu}_{\tau}$ measurement consistent with our Monte Carlo expectation. 
\begin{table}
\begin{tabular}{|c|c|c|}
\hline
Model & $a$ & $b$ \\
\hline
NuTeV \cite{Naples:2003ne} & 0.085 & 3.9\\
Kretzer et al. \cite{Kretzer:2002fr} & 0.080 & 5.6 \\
Hagiwara et al. \cite{Hagiwara:2003di} & 0.089 & 6.7 \\
Paschos et al. \cite{Paschos:2001np} & 0.105 & 4.1 \\
Nuance \cite{nuance} & 0.089 & 4.7 \\
\hline
\end{tabular}
\caption{The best fit parametrization for the various CC~$\nu_{\tau}$ cross-section models. See Equation~\ref{aandb} for definition of $a$ and $b$}\label{parameters}.
\end{table}


\setcounter{equation}{0}
\section{Experimental Setup And Event Simulation}
\label{sec:experiment}

\subsection{Liquid Argon Time Projection Chamber Technology}
The LAR20 detector provides a typical design for an ultra-large LArTPC~\cite{Baller}. We assume a 100~kt$\cdot$yr exposure for our event rate calculations, a modest estimate with the proposed 17~kt fiducial volume design and $>$10~years of data taking. 
Three wire planes, each with 3~mm wire spacing and 2~MHz sampling, will be instrumented. This granularity is insufficient to directly
reconstruct the tau decay kink and so the analysis presented
here relies on indirect methods of isolating CC~$\nu_\tau$/$\overline{\nu}_{\tau}$ interactions. The TPC's active volume will be $15\times 15\times 34$~m$^3$, featuring multiple field cages with voltages of 500 V/cm (corresponding to a drift velocity of 1.5~mm/$\mu$s).  Modest shielding is required for such a large detector so that the atmospheric neutrino events are not masked by and/or confused with cosmic
ray tracks and interactions. LAR20 is proposed for the 800~foot level of DUSEL, which
provides more than enough shielding for this analysis. Although a magnetized LArTPC has been demonstrated to work~\cite{magneticlartpc}, the current LAR20 design does not include this feature. It should be noted that a magnetic field would greatly enhance the analysis described here, allowing for possible $\nu_{\tau}$ and $\overline{\nu}_{\tau}$ separation and finer energy resolution, among other things. 

The spatial resolution for reconstructed tracks in the detector is at the millimeter scale in the drift and wire directions. A particle's energy can be reconstructed by adding up all of the charge collected along a stopping track. In the case that the particle is identified with some confidence, measuring the distance of range-out and/or multiple-scattering~\cite{multiplescattering} can also be used to reconstruct the energy of the particle. These techniques can be used simultaneously to measure the energy more precisely. Although the actual energy resolution is dependent on the particle's energy and identity, the energy resolution for LArTPCs is usually quoted at the few percent level~\cite{flare}. An example of data-based LArTPC neutrino event reconstruction can be found in Ref.~\cite{50liter}. In this analysis, we consider energy resolutions of 0\% and 15\%, reconstructed zenith angle resolutions of 0$^{\circ}$ and 10$^{\circ}$, and 100\% charged particle identification efficiency.

Although precise energy/angular resolution is not particularly vital to this measurement as demonstrated below, the identification of charged and neutral pions is. Especially at $>$1~GeV energies, separating a charged or neutral pion from a proton/electron/muon/kaon, is nearly 100\% efficient with LArTPCs. Particle identification proceeds in several ways. First, there is the combination of energy deposition per unit distance (d$E$/d$x$) and range. d$E$/d$x$ itself can distinguish highly ionizing particles such as protons and kaons from minimum ionizing ($\sim$2.1~MeV/cm in liquid argon) particles such as charged pions and muons with close to 100\% efficiency~\cite{2km}. High energy electrons and gammas (usually from neutral pion decay) are identified confidently as they lose most of their energy via Bremsstrahlung and $e^{+}/e^{-}$ pair production, creating a well defined electromagnetic shower. The experimentalist can also take advantage of the gamma's 18~cm conversion length in liquid argon, often resulting in a discernible gap between the interaction vertex (gamma creation point) and the beginning of the shower.  

Pions are separated from muons via hadronic multiple scattering (with hadronic interaction length of 84~cm), nuclear capture, and decay products. A particle that travels longer than a few hadronic interaction lengths without a secondary interaction can be identified as a muon. In the case that a negatively charged particle does not decay in flight, a $\pi^-$~($\mu^-$) will capture on argon 100\%~(76\%) of the time. Negatively charged pions and muons can be differentiated in this way as the nuclear capture products are significantly different in each case. The decay chains ($\pi\rightarrow\mu\rightarrow e$) and ($\mu\rightarrow e$) can also be used to separate pions from muons. 

The majority of atmospheric $\nu_{\tau}$/$\overline{\nu}_{\tau}$ interactions are high $Q^{2}$, deep-inelastic scattering events usually with large multiplicity and vertex activity. Such events can be difficult to fully reconstruct for Cherenkov-based experiments and detectors with weak spatial resolution. Disentangling long, energetic tracks (possibly overlapping rings in the case of a Cherenkov-based detector) is relatively simple for LArTPCs featuring three-dimensional imaging and millimeter resolution in a homogeneous and fully active volume. Even in the case of multiple $\pi^{0}$ production and subsequent decays ($\pi^{0}\rightarrow\gamma\gamma$), gamma pair matching is straightforward with the three-dimensional imaging capabilities of LArTPCs and the help of a reconstructed invariant $\pi^{0}$ mass. 

\subsection{Neutrino Event Generation}
\label{sec:evgen}
The Nuance (version 3) neutrino event generator~\cite{nuance} has been employed to simulate all-flavor atmospheric neutrino interactions on argon. The Nuance source code, originally created for simulating the interactions of atmospheric neutrinos with water, has been modified slightly in order to properly simulate the neutrino-argon interaction and the propagation of the resulting hadrons through the argon nucleus. The reader is referred to the Nuance documentation for detailed information on the various cross-section and nuclear process models used in the program. Modeling the energy and angular distributions of the products of intra-nuclear collisions is difficult as there are many types of interactions and little data. In Nuance, hadrons are stepped through an argon nucleus with measured radially-dependent density distribution and Fermi momentum. Some relevant argon-specific parameters are listed in Table~\ref{table2}. The pion-nucleon and nucleon-nucleon cross-sections and angular distributions are largely based on HERA data~\cite{hera}. The understanding of the cross-sections and angular distributions of relevant processes such as pion absorption, charge exchange, and elastic and inelastic scattering within the argon nucleus will be greatly improved by the ArgoNeuT~\cite{argoneut}, MicroBooNE~\cite{microboone}, and ICARUS T-600~\cite{icarus} experiments in the near future. These experiments will also improve our knowledge of the neutrino-on-argon cross-sections themselves. Similarly, MINER$\nu$A~\cite{minerva} will greatly enhance our knowledge of cross-sections and intra-nuclear interactions using multiple nuclear targets in the near future. The authors also look forward to a full intra-nuclear cascade simulation from the GENIE collaboration~\cite{genie}, featuring the simulation of ``all (intra-nuclear) reactions on all nuclei". 

Of special relevance to this paper, Nuance decays tau-leptons with the TAUOLA (version 2.6) package~\cite{tauola}. In CC quasi-elastic and deep-inelastic $\nu_{\tau}$/$\overline{\nu}_{\tau}$ interactions, the polarization of the tau is calculated using the appropriate form factors~\cite{tauola2}. The tau is considered completely polarized in resonant interactions, where the form factors are usually ambiguous and/or difficult to calculate~\cite{nuance}.

\begin{table}
 \begin{center}
\begin{tabular}{|c|c|}\hline
  Liquid argon& Nucleon binding energy = 29.5 MeV \ \\ 
           & Fermi momentum (p) = 242 MeV\\
           & Fermi momentum (n) = 259 MeV\\
           & Density = 1.396 g/cm$^{3}$\\
\hline
  Nuclear density & c = 3.53 fm  \\
  $\rho(r)=\frac{\rho_{0}}{1+e^{(r-c)/z}}$ & z = 0.542 fm \\
\hline
\end{tabular}
\end{center}
\caption{Some relevant parameters in the simulation of neutrino interactions on (liquid) argon.}\label{table2}
\end{table}

\section{Hadronic Tau Decays, Inclusive}
\label{sec:tau_to_x}

Identifying a $\nu_{\tau}$/$\overline{\nu}_{\tau}$ on an event-by-event basis will be extremely difficult for a detector unable to resolve the kink from the tau decay, occurring only a fraction of a millimeter from the event vertex for the most relevant atmospheric neutrino energies. However, statistically inferring the presence of CC~$\nu_{\tau}$/$\overline{\nu}_{\tau}$ interactions in a capable detector is possible by analyzing event kinematics. 

The tau decays to one or more hadrons and a neutrino about 65\% of the time, with the branching ratios of the various channels well known. For convenience, we tabulate the relevant decay modes in Table~\ref{table}. Looking for exclusive hadronic final states may assist in confirming the $\nu_{\tau}$/$\overline{\nu}_{\tau}$ appearance hypothesis. However, an exclusive analysis (requiring a specific pion final state) is complicated by an increased dependence on the simulation of the relatively uncertain final state interaction processes. We briefly consider the exclusive channels in the following section. In this section, we consider both inclusive pionic decays of $\nu_{\tau}$ and $\overline{\nu}_{\tau}$ together. Note that there is no reason that kaonic decays, compromising only about 3\% of the total hadronic channel, cannot be included in this sample. However, we do not consider them for simplicity and because of their negligible effect on the analysis.  

The largest background to $\nu_{\tau}$/$\overline{\nu}_{\tau}$ detection via hadronic-channel-inclusive pion decay kinematics is the atmospheric neutrino NC interaction involving the excitation(s) and then subsequent decay(s) of a nuclear resonance to a pion and a nucleon (with a small pion contribution from elsewhere, such as NC coherent production). CC~$\nu_{\mu}$/$\overline{\nu}_{\mu}$ and $\nu_{e}$/$\overline{\nu}_{e}$ events are considered a negligible background for $\nu_{\tau}$/$\overline{\nu}_{\tau}$ detection via the hadronic channels as the efficiency of GeV-scale electron and muon identification is near 100\% for LArTPC-based neutrino detectors~\cite{2km}. Separation of the CC and NC channels is critical to the detector's search for $\nu_{e}$/$\overline{\nu}_{e}$ appearance in the accelerator-based part of the experiment. 

Pions originating from neutrino-induced nuclear resonance decays (and elsewhere) and pions from tau decay form considerably different kinematic distributions. Variables such as visible energy, average energy per pion, angle between the highest energy pions, angle between the highest energy pion and the (reconstructed) initial neutrino angle, energy of the highest energy pion, invariant mass of the pionic system, and more are all useful, correlated, observables in differentiating pions originating from $\nu_{\tau}$/$\overline{\nu}_{\tau}$ events and pions from the NC background. 

For a 100~kt$\cdot$yr exposure, 77.2~CC~$\nu_{\tau}$+$\overline{\nu}_{\tau}$ events are expected, based on the previously discussed oscillated atmospheric neutrino flux and the Nuance cross section predictions. Of these, 46.3~events feature at least one charged pion from tau decay. CC~$\nu_{\tau}$/$\overline{\nu}_{\tau}$ events involving a charged lepton from the tau decay are not considered here, even if a non-tau-decay-pion escapes the nucleus. Visible energy ($E_{\mathrm{vis}}$), cos(reconstructed zenith angle) ($\mathrm{cos}(\theta_{\mathrm{zenith (rec.)}})$) and energy of the highest energy neutral or charged pion ($E_{\uparrow\pi}$) have been found to be the most useful variables in differentiating signal from background. These variables also happen to be among the simplest to measure experimentally. We initially consider events with at least one visible charged pion. Note that the pions in the CC~$\nu_{\tau}$+$\overline{\nu}_{\tau}$ sample are not all necessarily from the tau decay (e.g. they might be resonant-induced) although a tau decay to at least one charged pion is required to enter the sample.

Events with $\mathrm{cos}(\theta_{\mathrm{zenith (rec.)}})$$<$$-$0.2 and one or more charged pions enter the $\nu_{\tau}$+$\overline{\nu}_{\tau}$ candidate sample.  Reconstructed zenith angle is defined assuming full knowledge of all final-state and tau-daughter non-neutrino/neutron particles. The reconstructed zenith angle cut is applied to reduce the downward-going NC background as few $\nu_{\tau}$/$\overline{\nu}_{\tau}$ events are expected in this region. The cut is stricter than the usual $\mathrm{cos}(\theta_{\mathrm{zenith (rec.)}})$$<$0.0 cut in order to compensate for the limitations of reconstructed angle resolution in differentiating upward- from downward-going events. 

\begin{figure}
\includegraphics[width=8cm]{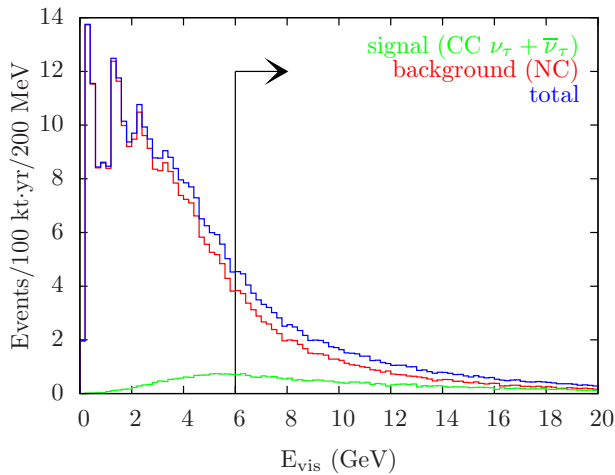}
\caption{The visible energy for NC interactions with at least one final state charged pion and CC~$\nu_{\tau}$+$\overline{\nu}_{\tau}$ interactions with the tau decaying to at least one charged pion. Events with $E_{\mathrm{vis}}$$>$6.0~GeV enter the final sample.}\label{visenergy}
\end{figure}

With a 3.5~GeV threshold, CC~$\nu_{\tau}$/$\overline{\nu}_{\tau}$ interactions generally occur at higher energies than the other possible atmospheric neutrino interactions. Visible energy is defined as the sum of the energy of all primary particles and tau decay products minus the energy contribution of any and all neutrinos and neutrons. Figure~\ref{visenergy} shows the visible energy of our sample, separated into NC, CC~$\nu_{\tau}$+$\overline{\nu}_{\tau}$, and total after simulating atmospheric neutrinos with energy 3$<$$E_{\nu}$$<$30~GeV. Requiring $E_{\mathrm{vis}}$$>$6.0~GeV is found to be effective in isolating the $\nu_{\tau}$/$\overline{\nu}_{\tau}$ events from the NC background. The visible energy cut was chosen to improve signal-to-background, although a thorough optimization of this cut has not been performed. Ref.~\cite{icanoe}~also found a 6-7~GeV visible energy cut useful for isolating atmospheric $\nu_{\tau}$/$\overline{\nu}_{\tau}$ events.  

\begin{figure}
\includegraphics[width=8cm]{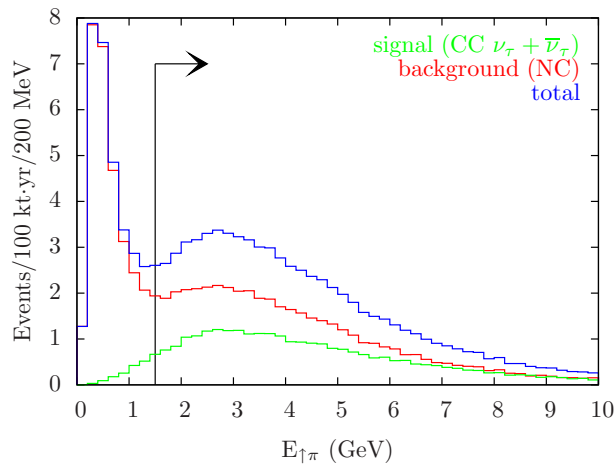}
\caption{The energy of the highest energy pion, neutral or charged, for NC interactions with at least one final state charged pion and CC~$\nu_{\tau}$+$\overline{\nu}_{\tau}$ interactions with the tau decaying to at least one charged pion. Non-tau-decay-pions (e.g. resonant-induced) are also considered in the CC~$\nu_{\tau}$+$\overline{\nu}_{\tau}$ sample. Events with $E_{\uparrow\pi}$$>$1.5~GeV enter the final sample.}\label{highpion}
\end{figure}

Figure~\ref{highpion} shows $E_{\uparrow\pi}$ for each sample after the visible energy cut. After further requiring $E_{\uparrow\pi}$$>$1.5~GeV, we are left with 73.4~total events over a background of 44.8~events, a 4.3$\sigma$ excess (with Poisson statistics only) assuming experimental data that is consistent with the Monte Carlo expectation. This excess has been found to be largely insensitive to the effects of even modest energy and reconstructed angle resolutions. Applying a Gaussian smear (with a 15\% sigma) to each event's $E_{\mathrm{vis}}$ and $E_{\uparrow\pi}$ and a Gaussian smear (with a $\mathrm{10}^{\circ}$ sigma) to $\mathrm{cos}(\theta_{\mathrm{zenith (rec.)}})$ (all smears together), left the excess at 3.9$\sigma$. Similarly, varying each cut individually by $\pm$15\%, while leaving the other two cuts at their nominal values, dropped the excess down to 4.2$\sigma$ at worst. 

The background event rates quoted above rely on the understanding of the NC cross sections in the $\sim$few to 30~GeV energy range and final state interactions. Fortunately, experiments such as those mentioned in Section~\ref{sec:evgen} will soon improve our knowledge of these processes. The Nuance event generator represents our best estimate of the NC background and we employ it with the understanding that the ``real" background (and signal) may be substantially different from that which is predicted. The apparent flexibility of the resolution/cuts described above demonstrates that the sensitivity quoted will only minimally depend on the modeling of NC cross sections and final state interactions. Moreover, a strong excess remains even in the case that Nuance severely under-predicts the background. Arbitrarily increasing the background that passes all cuts by 50\% (while leaving the signal at its nominal level) leaves the excess at 3.5$\sigma$.
 
This analysis would benefit from a neural network and/or likelihood analysis involving the many correlated kinematic variables useful for discriminating CC~$\nu_{\tau}$/$\overline{\nu}_{\tau}$ from the NC background. Although a complicated multi-layer approach to inferring atmospheric $\nu_{\tau}$/$\overline{\nu}_{\tau}$ appearance could be utilized to improve sensitivity, we have shown that simple observables (and a finely-grained and efficient detector with particle identification capabilities) are all that is necessary for this measurement. 

\begin{table}
\begin{tabular}{|c|c|}\hline
$\tau^-$ decay mode & \ Branching ratio \ \\  \hline
$e^-\nu_{\tau}\overline{\nu}_e$ & \ 17.8\% \ \\ \hline
$\mu^-\nu_{\tau}\overline{\nu}_{\mu}$ & \ 17.4\% \ \\ \hline
$\pi^-\nu_{\tau}$ & \ 11.1\% \ \\  \hline
$K^-\nu_{\tau}$ & \ 0.69\% \ \\  \hline
$\pi^-\pi^0\nu_{\tau}$ & \ 25.4\%$^{\dagger}$ \ \\  \hline
$(\ge 1 h^-)(\ge 0 h^0)\nu_{\tau}$ & \ 62.6\%$^{\ddagger}$ \ \\  \hline
\end{tabular}
\medskip \\
\vspace{0.0cm}
\caption{Branching ratios of selected $\tau^{-}$ decay modes~\cite{Amsler:2008zzb}. $^\dagger$All but a small fraction ($0.3\%$) of this branching ratio is mediated by a $\rho^-$ resonance. $^{\ddagger}h^{-,0}$ is either a pion or a kaon.}\label{table}
\end{table}

\setcounter{equation}{0}
\section{Hadronic Tau Decays, Exclusive}
\label{sec:tau_to_n_pi}
\begin{figure}
\includegraphics{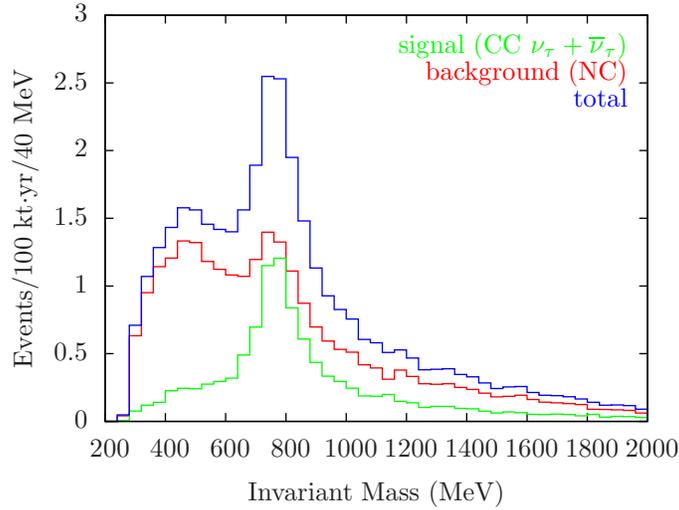}
\caption{The invariant mass of the highest energy charged and neutral pion for NC interactions and CC~$\nu_{\tau}$+$\overline{\nu}_{\tau}$ interactions with one neutral pion and at least one charged pion. Non-tau-decay-pions are also considered in the CC~$\nu_{\tau}$+$\overline{\nu}_{\tau}$ sample.}\label{invmass}
\end{figure}

The inclusive hadronic decays are broken up into their constituent exclusive channels in Table~\ref{table3}. The three cuts discussed above ($E_{\mathrm{vis}}$$>$6.0~GeV, $E_{\uparrow\pi}$$>$1.5~GeV, and $\mathrm{cos}(\theta_{\mathrm{zenith (rec.)}})$$<$$-$0.2) are used to arrive at the expected number of events. The categorization of the inclusive channels cannot simply be made according to the expected tau decay products and branching ratios as at least one non-tau-decay-pion appears in 68\% of the CC~$\nu_{\tau}$/$\overline{\nu}_{\tau}$ events simulated. For example, requiring the invariant mass of a charged and neutral pion pair to match the $\rho$ meson peak, in an effort to isolate the most likely decay channel ($\tau\rightarrow\pi^{\pm}\pi^0\nu_{\tau}$, with a branching ratio of 25.4\%), is complicated by these non-tau-decay-pions. Figure~\ref{invmass} shows the invariant mass of the highest energy charged and neutral pion in events with $>$0 charged pions, 1 neutral pion, $E_{\mathrm{vis}}$$>$6.0~GeV, and $\mathrm{cos}(\theta_{\mathrm{zenith (rec.)}})$$<$$-$0.2. Although a clear $\rho$ peak is visible, signal events often do not reconstruct well to the $\rho$ mass. Furthermore, the NC background contains a significant amount of $\rho$ resonances, a consequence of vector meson dominance. Requiring the highest energy charged and neutral pion invariant mass to be greater than 500~MeV along with the same three cuts on events with a single neutral pion and one or more charged pions only negligibly increases the sensitivity.      

Although the actual sensitivity to $\nu_{\tau}$/$\overline{\nu}_{\tau}$ appearance cannot be improved via an exclusive analysis, such an analysis can help to confirm that a possible excess is in fact from $\nu_{\tau}$/$\overline{\nu}_{\tau}$ interactions rather than non-standard neutrino interactions. That is, seeing the expected excess in each charged/neutral pion category consistent with tau decay (and the expected non-tau-decay-pion contribution) can rule out non-standard neutrino interactions as the source of an excess.

It is worth noting that atmospheric CC~$\nu_{\tau}$/$\overline{\nu}_{\tau}$ events with a kaon in the final state (i.e. $\tau\rightarrow K \nu$) should not be considered a background for the proton decay channel $p\rightarrow K^{+}\overline{\nu}$. An argon-bound-proton's decay kaon is expected to have a momentum of less than 500 MeV/c as it exits the nucleus, taking into account the effects of kaon rescattering (with or without an argon spectral function)~\cite{protondecay}. Conservatively requiring the kaon's momentum to be $<$500~MeV/c, less than 0.05 CC~$\nu_{\tau}$/$\overline{\nu}_{\tau}$ ($\tau\rightarrow K \nu$) events/100~kt$\cdot$yr are expected. 

\begin{table}
\begin{center}
\begin{tabular}{|c|c|c|}\hline 
Exclusive channel ($\pi$'s) & \ Total/Background & \ Excess significance ($\sigma$) \ \\  \hline
1 charged, 0 neutral & \ 2.3/1.6 & \ 0.6 \ \\ \hline 
1 charged, 1 neutral & \ 5.9/3.2 & \ 1.5 \ \\ \hline 
1 charged, $>$1 neutral & \ 6.9/3.5 & \ 1.8 \ \\ \hline
$>$1 charged, 0 neutral & \ 11.5/7.6 & \ 1.4 \ \\ \hline 
$>$1 charged, 1 neutral & \ 21.5/14.1 & \ 2.0 \ \\ \hline 
$>$1 charged, $>$1 neutral & \ 25.3/14.8 & \ 2.7 \ \\ \hline 
Total & \ 73.4/44.8 & \ 4.3 \ \\ \hline 
\end{tabular}
\end{center}
\caption{The expected number of CC~$\nu_\tau$+$\overline{\nu}_{\tau}$ and NC background events/100~kt$\cdot$yr after cuts for hadronic tau decay channels categorized by number of charged and neutral pions.}\label{table3}
\end{table}

\begin{figure}
\includegraphics{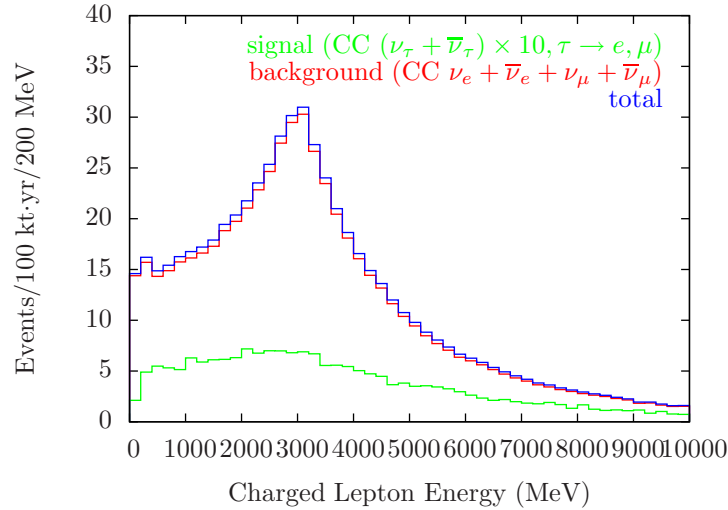}
\caption{The visible energy distribution of charged leptons from signal CC~$\nu_{\tau}$+$\overline{\nu}_{\tau}$ and background CC~$\nu_{\mu}$+$\overline{\nu}_{\mu}$+$\nu_{e}$+$\overline{\nu}_{e}$ events.}\label{lepton}
\end{figure}


\setcounter{equation}{0}
\section{Leptonic Tau Decays}
\label{sec:tau_to_ell}

The tau decays to a charged lepton and two neutrinos with a branching ratio of about 35\%. Accelerator-based $\nu_{\tau}$/$\overline{\nu}_{\tau}$ appearance experiments without the ability to see the decay kink in such events can infer the presence of a $\nu_{\tau}$/$\overline{\nu}_{\tau}$ interaction by searching for missing transverse energy (carried away by the two unseen neutrinos). This missing transverse energy is absent for background CC~$\nu_{\mu}$/$\overline{\nu}_{\mu}$/$\nu_{e}$/$\overline{\nu}_{e}$ events which don't feature final state neutrinos. However, an atmospheric $\nu_{\tau}$ appearance search is not afforded the luxury of a known beam direction and a missing transverse energy search is therefore difficult. 

Compared to the inclusive hadronic modes discussed above, the CC~$\nu_{\tau}$/$\overline{\nu}_{\tau}$ leptonic channels have a higher background and smaller signal. The leptonic channel's background (CC~$\nu_{\mu}$/$\overline{\nu}_{\mu}$/$\nu_{e}$/$\overline{\nu}_{e}$) cross-section is almost three times higher than its NC counterpart and the signal tau decays to a charged lepton about half as often as it does to one or more hadrons. Furthermore, there are fewer cuts available to separate background from signal. The powerful visible energy cut used for the inclusive hadronic final states above is rendered mostly useless for the leptonic case as the CC~$\nu_{\tau}$/$\overline{\nu}_{\tau}$'s unseen neutrino daughters push the event's visible energy down, overlapping more with the background CC~$\nu_{\mu}$/$\overline{\nu}_{\mu}$/$\nu_{e}$/$\overline{\nu}_{e}$ visible energy. Requiring $\mathrm{cos}(\theta_{\mathrm{zenith (rec.)}})$$<$$-$0.2 and $E_{\mathrm{vis}}$$>$3.0~GeV, we find 20 CC~$\nu_{\tau}$+$\overline{\nu}_{\tau}$ events with an expected background of 610 CC~$\nu_{\mu}$+$\overline{\nu}_{\mu}$+$\nu_{e}$+$\overline{\nu}_{e}$ events (see Figure~\ref{lepton}). Cuts involving the charged lepton's transverse momentum and direction, both with respect to the reconstructed neutrino direction, have been found to be largely ineffective. We find the leptonic channel an unlikely source for atmospheric $\nu_{\tau}$/$\overline{\nu}_{\tau}$ investigation.

\section{Conclusions} 
\label{sec:conclusion}
The $\nu_{\tau}$/$\overline{\nu}_{\tau}$ is the least well understood observed Standard Model particle with only ten events ever identified. Simply measuring the cross-section and testing the Standard Model prediction, after obtaining a sizable sample of events, will improve our understanding of this elusive particle. The prospect of measuring the atmospheric neutrino mixing parameters from a $\nu_{\mu}\rightarrow\nu_{\tau}$ ($\overline{\nu}_{\mu}\rightarrow\overline{\nu}_{\tau}$) appearance experiment, however imprecisely, rather than a $\nu_{\mu}\nrightarrow\nu_{\mu}$ ($\overline{\nu}_{\mu}\nrightarrow\overline{\nu}_{\mu}$) disappearance experiment would also be a strong corroboration of the three neutrino mixing model in general.

Along with offering an attractive detector for high- and low-energy accelerator-based neutrino oscillation experiments, atmospheric $\nu_{\mu}$/$\overline{\nu}_{\mu}$ disappearance measurements, proton decay searches, and sensitivity to supernova burst and diffuse neutrinos, among other things, a kiloton-scale LArTPC will also be capable of observing a significant number of atmospheric $\nu_{\tau}$/$\overline{\nu}_{\tau}$ appearance events. Contrary to naive expectations, we find that the most promising way of detecting atmospheric $\nu_{\tau}$/$\overline{\nu}_{\tau}$s is through the study of the tau hadronic decay modes as it is difficult to separate tau-decay leptons and leptons coming from CC~$\nu_{e}$/$\overline{\nu}_{e}$ and $\nu_{\mu}$/$\overline{\nu}_{\mu}$  interactions. After simple cuts on visible energy, reconstructed zenith angle, and energy of the highest energy pion in atmospheric neutrino events with no charged lepton, 28.5~tau neutrino events over a background of 44.8~events are expected from a 100~kt$\cdot$yr exposure, corresponding to a 4.3$\sigma$ excess signal with Poisson statistics only. Although the detection of such events might be considered ``indirect", the expected signal-to-background ratio over most of the kinematic range (after cuts) is $>$0.5 and even $>$1 at high energies (see Figure~\ref{highpion}). Furthermore, by studying the exclusive decay channels categorized by number of charged/neutrino pions, the possibility of an excess in the inclusive analysis being a result of some form of non-standard interaction, rather than $\nu_{\tau}$/$\overline{\nu}_{\tau}$ production, can be disfavored. 

\begin{figure}
\includegraphics[width=8cm]{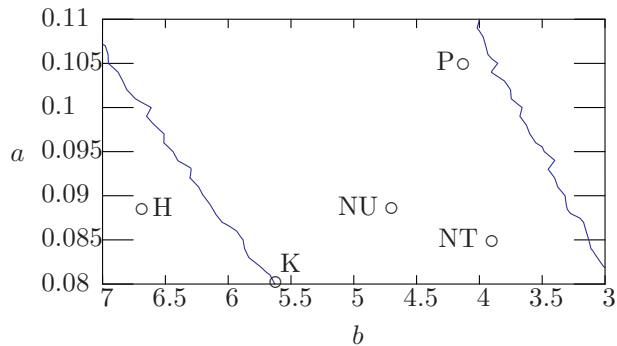}
\vspace{0.0cm}
\caption{The CC~$\nu_{\tau}$ cross-section parametrization sensitivity (see Equation~\ref{aandb}), assuming a measurement consistent with the Nuance prediction for neutrino-on-argon scattering. The lines denote the 1-$\sigma$ (statistical-error-only) allowed region. The other points denote models by Hagiwara et al. (H), Kretzer et al. (K), Paschos et al. (P) and the NuTeV collaboration (NT) for neutrino-on-isoscalar-target scattering. Note that the confidence level drawn is optimistic as it assumes full knowledge of the $\nu_{\tau}$:$\overline{\nu}_{\tau}$ ratio.}
\label{contour:one}
\end{figure}

\begin{figure}
\includegraphics[width=12cm]{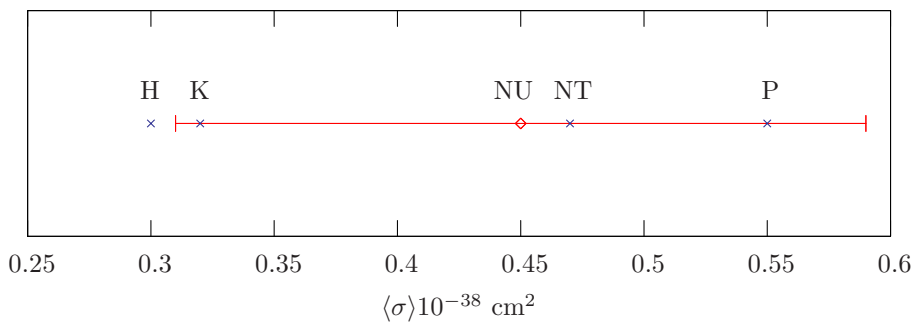}
\caption{The flux-averaged total CC~$\nu_{\tau}$+$\overline{\nu}_{\tau}$ cross-section (6$<$$E_{\mathrm{vis}}$$<$30~GeV), assuming a measurement consistent with the Nuance neutrino-on-argon prediction (with 1-$\sigma$ statistical error). 
We show the cross-section for neutrino-on-isoscalar-target from other models for reference. 
Letters adjoining the dots refer to the same models as in Figure~\ref{contour:one}.
}
\label{contour:two}
\end{figure}

Given a measurement consistent with the simulated sample, the expected statistical-error-only sensitivities to the CC~$\nu_{\tau}$ cross-section parametrization and the flux-averaged total CC~$\nu_{\tau}$+$\overline{\nu}_{\tau}$ cross-section (6$<$$E_{\mathrm{vis}}$$<$30~GeV) from a 100~kt$\cdot$yr exposure are shown in Figures~\ref{contour:one} and~\ref{contour:two} along with the predictions of various cross section models. Such measurements would provide basic tests of the Standard Model and vastly improve our knowledge of the elusive third neutrino weak eigenstate.
 

%
%

\section*{Acknowledgements} 
We would like to thank Bonnie Fleming, Mitchell Soderberg, Chris Walter, and Geralyn Zeller for helpful discussions. We thank the hospitality of Fermilab where a lot of this work was carried out. AdG and SS are sponsored in part by the US Department of Energy Contract DE-FG02-91ER40684. JC is sponsored in part by the National Science Foundation grant number PHY0847843.

\bibliography{atm_nu_tau12}
\bibliographystyle{h-physrev}
\end{document}